# A Quantitative Information Measure applied to Texture Perception Attributes during Mastication

Leen Sturtewagen[1], Harald van Mil[2], Marine Devezeaux de Lavergne[3,4], Markus Stieger[3,4], Erik van der Linden[1], Theo Odijk[5]

[1] Laboratory of Physics and Physical Chemistry of Foods, Wageningen University, Bornse Weilanden 9, 6708 WG Wageningen, The Netherlands
[2] Mathematical Institute, Leiden University, Niels Bohrweg 1, 2333 CA Leiden, the Netherlands
[3] TI Food and Nutrition, Nieuwe Kanaal 9A, 6709PA Wageningen, The Netherlands
[4] Division of Human Nutrition, Wageningen University, Stippeneng 4, 6708WE Wageningen, The Netherlands
[5] Lorentz Institute for Theoretical Physics, Niels Bohrweg 2, 2333 CA Leiden, The Netherlands

## Abstract

We have calculated a quantitative measure of information of experimentally determined temporal dominance of sensations (TDS) frequencies of texture attributes, for a set of diverse samples throughout the mastication cycle. The samples were emulsion filled gels, two-layered emulsion filled gels, and sausages. For the majority of the samples we find one master curve, where swallowing takes place after the information increases from its minimum. The master curve may indicate a simplifying principle during mastication and subsequent swallowing.

We have also calculated a particular complexity measure. This measure displays an increase just before swallowing.

## 1. Introduction

Texture perception resulting from food consumption is partly a result of the combination in our brain of electrical signals that arrive from our five sense organs via the nerves (Rolls, 2005; Rolls et al., 2003; Verhagen and Engelen, 2006). The combination of signals is exemplified by the existence of texture-taste interactions (Burns and Noble, 1985) and texture-aroma interactions (Bult et al., 2007; Saint-Eve et al., 2004). For another part this perception is dependent on previous experiences (Mojet and Köster, 2005), mood (Gibson, 2006), eating behavior (Devezeaux de Lavergne et al., 2015a), social setting (Cardello et al., 2000; King et al., 2004), location (Edwards et al., 2003; Stroebele and De Castro, 2004), etc… The difficulty that is involved in the integration of all of these aspects makes understanding texture perception resulting from food consumption a challenging task.

For texture perception, an important point is the time of swallowing. With respect to this point in time, two material properties that have been pointed out are the degree of structure and the lubrication of the food material (Hutchings and Lillford, 1988). Because different food materials exhibit a different degree of structure and a different degree of lubrication upon swallowing (Hutchings and Lillford, 1988), it has not yet been possible to develop one quantitative model for different food materials that describes the degree of structure and the lubrication during mastication until the moment of swallowing.

Besides this specific time of swallowing, it is important to know the texture perception over the entire mastication time. The past decades have shown developments in this regard. The first step consisted of making a division between visual assessment, first bite, early and late mastication, swallowing and residual properties (Brandt et al., 1963). In this method (texture profile), each panel member needs to integrate the perception of each texture attribute over time to a single intensity value. The next step consisted of quantifying the temporal response of one texture related attribute (Larson-Powers and Pangborn, 1978). This is referred to as time intensity measurements. If one is interested in more than one attribute, the usual approach has been to repeat measuring time intensity profiles for each attribute (Guinard et al., 2002). Lately, temporal dominance of sensations (TDS) has been applied, a method in which for each moment in time the most dominant attribute is chosen (Pineau et al., 2009).

This TDS method yields the frequency of the most dominant attribute as a function of time. The time is usually being normalized by the time duration between food intake to swallowing. For one type of food, depicting a variety of attributes, TDS yields a spectrum of attribute frequencies over time (Di Monaco et al., 2014; Pineau et al., 2009). One usually presents the data after averaging over all panel members.

The choice by a panel member for the dominant attribute is made from a number of predefined attributes. Once a choice has been made, information is in fact conveyed from the panel member to the outside world. In other words, information, about what the panel member senses, has increased, in the outside world. Accordingly, uncertainty about what the panel member dominantly perceives has decreased, i.e. from the initial uncertainty regarding the set of possible attributes to the one dominant attribute that was finally chosen. Interestingly, one has a quantitative measure for this information, and its counterpart, uncertainty, or entropy (Rothstein, 1951; Shannon, 1948; Szilard, 1929). It is the purpose of this article to explore the value of using these quantitative measures in analyzing sensory perception. Hereto we calculate these quantitative measures for a specific set of TDS-data (Devezeaux de Lavergne et al., 2016, 2015b, 2015a).

Such quantitative measures have not been calculated from TDS data, but in sensory science, the concept of information has been used, for example, in experiments on absolute judgement (Garner and Hake, 1951; Miller, 1953, 1956), and on reaction times (Hick, 1952). It was also used for quantifying the maximal capacity of a person to perceive something and to process the information (Attneave, 1954; Miller, 1956; Munsinger and Kessen, 1964). A very recent example of the use of information can be found in the field of consciousness and awareness (Guevara Erra et al., 2016), who identified "features of brain organization that are optimal for sensory perception". They suggested that "consciousness could be the result of an optimization of information processing".

In order to further explore possibly underlying principles leading to the TDS profiles we have calculated a particular complexity measure. We note that also the concept complexity is known in the field of sensory science. However, quantifying complexity has not been a major effort. This may be partly caused by the fact that complexity is still an ill-defined concept within sensory science. For example, Berlyne (1960) characterizes complexity as a multidimensional concept where something becomes more complex when there are more interacting distinguishable elements. This resonates with many authors who suppose a linear relationship between complexity and the amount of stimuli or compounds (Olabi et al., 2015; Porcherot and Issanchou, 1998; Reverdy et al., 2010; Ruijschop et al., 2010; Weijzen et al., 2008). In addition to these semi-quantitative measures, at the same time only rather qualitative measures for complexity have been used, like "the difficulty to describe or identify the stimulus" (Olabi et al., 2015; Porcherot and Issanchou, 1998; Ruijschop et al., 2010) and "the inverse of simplicity" (Jellinek, 1990; Mielby et al., 2014, 2013, 2012; Soerensen et al., 2015).

In this article we start with discussing our results on information and entropy, and then discuss our results on complexity.

# 2. Materials and methods

## 2.1 Temporal Dominance of Sensations (TDS) data and pre-processing

For our analysis we used previously published data of TDS studies on various samples with only texture attributes included for the TDS experiment. The samples that were investigated on their sensory perception have been described elsewhere in more detail, for further details about the experimental method and panel size, the reader is directed to (Devezeaux de Lavergne et al., 2016, 2015b, 2015a). Three sets of samples were studied. The first two sets of samples are model gel food systems. The first set are 8 emulsion filled gels with different mechanical properties (Devezeaux de Lavergne et al., 2015b). The gels varied in fracture stress (low and high), fracture strain (low and high) and in the emulsifier used (WPI or Tween 20). The second set are 10 emulsion filled gels with mechanical contrast (Devezeaux de Lavergne et al., 2016). They consist of two layers with different mechanical properties (low or high gelatin or agar concentration). The third set of samples were 2 different kind of sausages: a hard (Ardenner) sausage and a soft (Berliner) sausage (Devezeaux de Lavergne et al., 2015a).

The sensory perception was determined for a variety of attributes by means of panels. The panels that did the experiments with the model foods were selected based on their "discriminative abilities" for the different textures. They had extensive experience with sensory experiments with semi-solid model foods and Qualitative Descriptive Analysis (QDA) (Devezeaux de Lavergne et al., 2016, 2015b). The panels that did the experiment with the sausages were consumers that were selected based on their eating behavior, in particular their eating duration. They had less experience with sensory experiments (Devezeaux de Lavergne et al., 2015a).

During the TDS experiment a number of panelists, $n_p$, were asked to select the most dominant attribute during mastication of a sample from a set that contains a pre-defined number of attributes, $N_a$. The dominance of an attribute was defined as "the attribute that attracts the most attention at a given time point". The experiment is set up in such a way that only one dominant attribute can be selected by a panelist at any given time point during mastication. The selected attribute is considered dominant until the next attribute is selected. The obtained attribute sequence is registered as a binary response, data are coded as "0" if the attribute, $a$, is not dominant for a given point in time and "1" if it is dominant. For each measurement and each panelist, the time starts when they place the sample in their mouth (defined as $t = 0$) and stops when they swallow the sample (defined as $t = 100$). Each panelist repeats their selection of the dominant attribute for each type of sample as a function of time, with a number of repetitions, $n_r$. For each measurement, a panelist may need another mastication time. We normalize the mastication time by the total mastication time of each panelist in order to obtain normalized time from 0 to 100. The samples were presented to the panel in a balanced randomized design. We define the probability of attribute selection by the panel at, at a time, $t$, $p(a|t)$. Taking the total number of times that an attribute $a$ is chosen by the panel at a time $t$ as $n_a$, we have $p(a|t) = n_a/(n_r * n_p)$. Analyses were conducted in R 3.2.5 (R Core Team, 2016).

## 2.2 Information, uncertainty and entropy

We refer to the Appendix for some background to the quantification of the terms information, uncertainty, and entropy. In short, information on a system is what you know about the system while uncertainty and entropy are what you do not know about a system. Therefore, information equals minus entropy.

Applying the expression for the total entropy, $H(t)$, in our case in the selection by the panel of an attribute $a$ at a given time $t$, which is related to the attribute selection probability of $p(a|t)$, we have:

$$H(t) = -K \sum p(a|t) \log_2 p(a|t), \qquad Eq.\,1$$

in which the normalization factor $K = 1/\log_2 N_a$, with $\log_2$ denoting the logarithm with base 2, and in which $\Sigma$ denotes the summation over all attributes $a$ in the set.

Based on this equation, when the probability that an attribute is selected at a certain time is equal to 1, the uncertainty (entropy) is 0. In this case, there is only one attribute perceived as dominant for the whole panel, so there is no uncertainty about the dominant attribute and a maximum of information is gained from the selection of the attribute. However, when the probability for the selection of all attributes is $1/N_a$, the uncertainty with respect to the selection of an attribute becomes 1. In this case, all attributes have an equal probability to be selected by the panel, so the uncertainty for selecting any attribute becomes maximal and the information gained from the selection of an attribute becomes minimal.

In the beginning of mastication, the total number of selected attributes, $n_a$ remains 0 for a short while. It takes a certain time (lag time) for a panelist to select a dominant attribute. This lag time is different for each panelist. With continuing mastication, panelists start to perceive the attributes and select their first most dominant one. After a while all panelists have selected a dominant attribute. Because of this lag time in selecting the first dominant attribute a so-called estimator is used. This estimator, the Chao-Shen estimator, accounts for the fewer selected attributes in the beginning of mastication and ensures that the uncertainty is not over- or under-estimated (Chao and Shen, 2003; Hausser and Strimmer, 2008). The Chao-Shen estimator is defined by

$$\widehat{H(t)} = -K \sum \frac{\widehat{p(a|t)} \log_2 \widehat{p(a|t)}}{\left(1 - \left(1 - \widehat{p(a|t)}\right)^{n_a}\right)} \qquad Eq.\,2$$

In which $\widehat{p(a|t)} = \left(1 - \frac{m}{n_a}\right) p(a|t)$, with $m$ the number of attributes that is selected only once over all $n_r * n_p$ measurements, and $n_a$ is the total amount of selected attributes at time $t$. We use this measure $\widehat{H(t)}$ for the uncertainty, or entropy, regarding attribute selection by the panel, as a function of time.

# 3. Results and discussion

**3.1 Information, uncertainty, entropy and TDS data**

Figure 1 shows the entropy over time curve for one sample. It shows how the entropy for the perceived texture of the sample changes over mastication time, which has been normalized by the total mastication time. In the beginning of mastication time, the entropy (uncertainty) is low: the panel is just starting to perceive the sample and they are selecting their first dominant attribute. The information gained is high. With continued mastication, the sample is broken down and mixed with saliva. This alters the structure of the sample and thus the perceived texture. Upon further mastication, the food is broken down even further, and a swallowable bolus is formed during this process. The information gained about the most dominantly perceived texture attribute decreases and reaches a minimum about the middle of the mastication process. The panel is not in consensus anymore about the most dominant texture attribute at that time. Many attributes get selected. The entropy (uncertainty) is maximal. The panel subsequently reaches more consensus about the most dominant attribute as can be seen from the subsequent increasing information gain and entropy loss. At the moment of swallowing the entropy is about 0.5. We note that this implies that at the point of swallowing, more than one attribute is perceived as dominant. This in turn implies that there is not one specific attribute responsible for triggering the swallowing action.

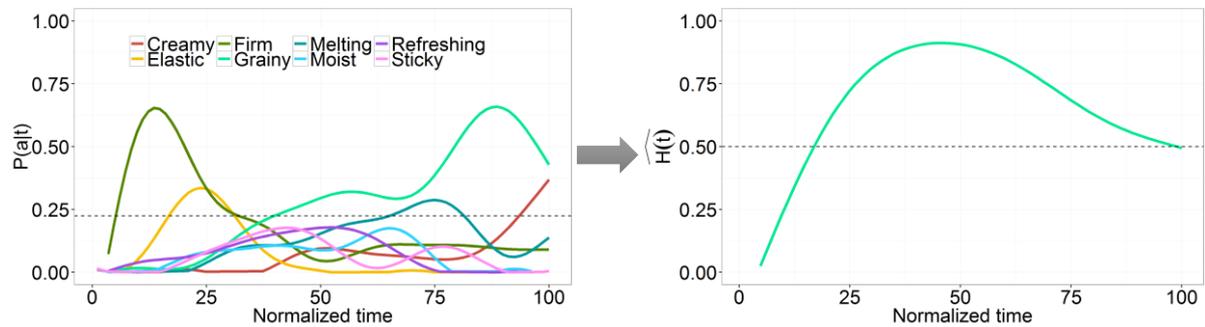

Figure 1: (Left) Temporal Dominance of Sensations plot of one sample of emulsion filled gels (the sample had high engineered fracture stress ($\sigma$) and low engineered fracture strain ($\epsilon$) and WPI as an emulsifier). (Right) Temporal entropy curve for the same sample. Time is normalized as a % of total mastication time per measurement (n = 10, triplicate). TDS data from Devezeaux de Lavergne et al. (2015b).

Figure 2 shows the temporal entropy curves for a variety of samples with different mechanical properties. The samples varied in fracture stress (low and high), fracture strain (low and high) and in the emulsifier used (WPI or Tween 20). Even though the mechanical properties of the samples were highly different, we see that the (normalized) entropy curves follow the same curve. At the start of mastication time, the entropy (uncertainty) is low, there is consensus about the most dominant attribute. In the middle of mastication time, the information reaches a minimum, there is no consensus about the most dominant attribute and many different attributes are being selected randomly by the panel at that time. Towards the end of mastication, the panel reaches more consensus about the most dominant attribute and the entropy (uncertainty) decreases again, but not to zero. Again, apparently more than one attribute is being perceived at the start of the swallowing process and there is not one attribute responsible for triggering the swallowing action. For two gels ($H\sigma L\epsilon T$ and $L\sigma L\epsilon T$), the uncertainty about the most dominant texture attribute is still quite high at the point of swallowing. Both these gels have a low fracture strain and their oil droplets where emulsified with Tween 20. This means that they are brittle gels and their oil droplets are not bound to the matrix. Because of the brittleness, there is a high probability that they are perceived as grainy and less as creamy. The brittleness apparently leads to a higher uncertainty in selection of an attribute.

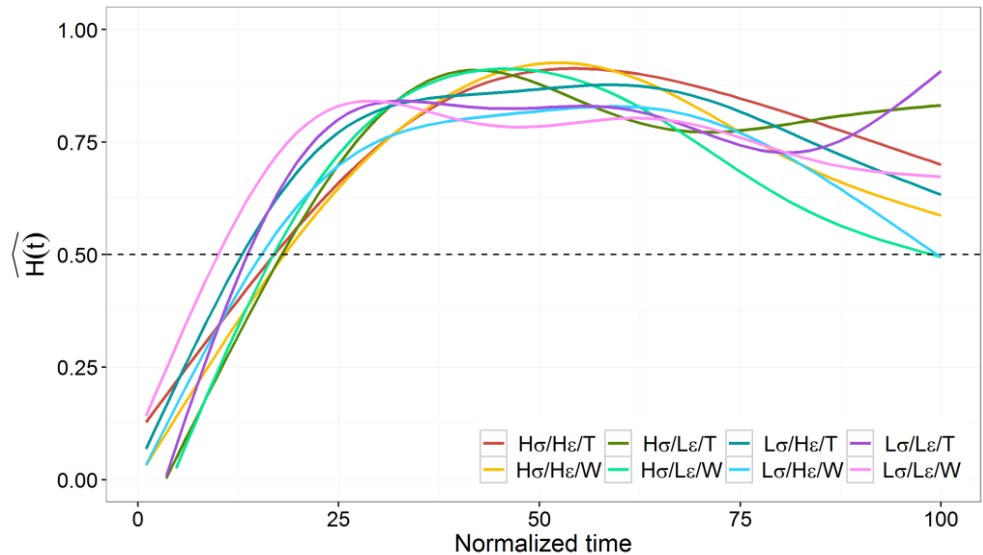

Figure 2: Temporal entropy curves of emulsion-filled gels (with low (L) or high (H) engineered fracture stress ($\sigma$) and strain ($\epsilon$) and WPI (W) or Tween 20 (T) as an emulsifier). Time was normalized as a % of total mastication time per measurement (n = 10, triplicate). TDS data from Devezeaux de Lavergne et al. (2015b).

Figure 3 shows the temporal entropy curves for emulsion filled gels. These gels exhibited a mechanical contrast. These samples consist of two layers with different mechanical properties (low or high gelatin or agar concentration). The entropy follows the same curve as the samples in Figure 2. One sample (HA + HA) shows a lower entropy than the others. This sample contains two of the same layers with a high agar concentration. This makes the sample first firm and then grainy. During mastication, there is a high probability it is being perceived as grainy. This reduces the uncertainty about the perceived texture. We come back to this point in more detail below.

For this set of samples the panel was more experienced with the TDS technique. Panelists that are more experienced with the TDS technique tend to select their first attribute faster than panels with little experience. Because the panel selected their first attribute faster (after 2.0 ± 1.3 seconds on average compared to 3.4 ± 2.3 seconds on average for the less experienced panel), the uncertainty rises more steeply in the beginning and reaches a longer plateau in the middle. This can be seen in Figure 3. The longer plateau exhibits a fluctuation in information in the middle of the mastication process, which we attribute to the fact that the gels exhibit a mechanical contrast.

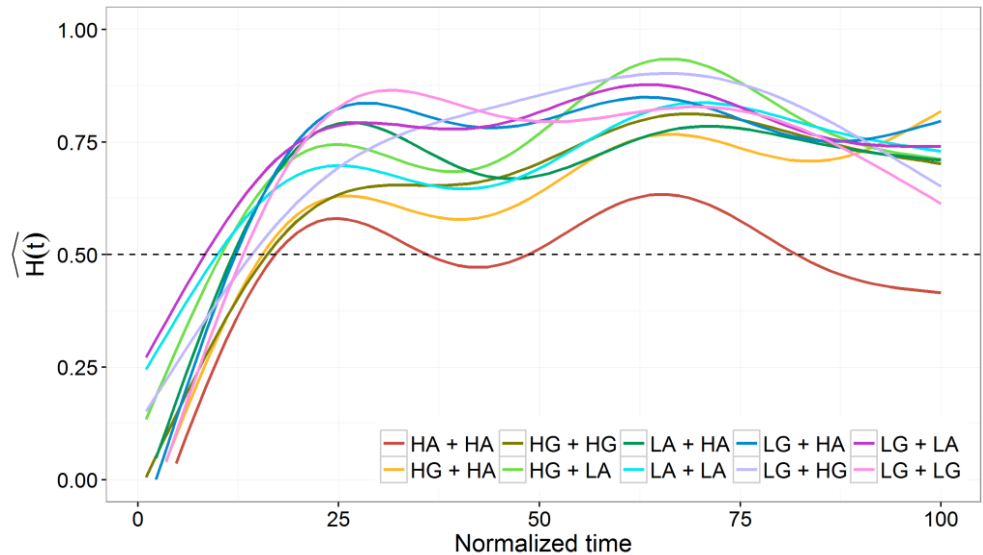

**Figure 3:** Temporal entropy curves of emulsion filled layered gels with mechanical contrast (low (L) or high (H) gelatin (G) or agar (A) concentration). Time was normalized as a % of total mastication time per measurement (n = 10, triplicate). TDS data from Devezeaux de Lavergne et al. (Devezeaux de Lavergne et al., 2016).

In Figure 4 the temporal entropy is compared for two different panels while consuming sausages. The panels were selected based on their eating behavior: a panel with a long mastication time (40.8 ± 9.7 seconds) and a panel with a short mastication time (18.2 ± 3.9 seconds). The panels did the TDS experiment for two different sausages: a hard (Ardenner) sausage and a soft (Berliner) sausage. Even though the sausages where different in mechanical properties, we see that the uncertainty about the most dominant texture attribute for the slow eaters follows a similar pattern as in Figure 2 and 03. In contrast, the fast eaters maintain a high information/low uncertainty upon swallowing.

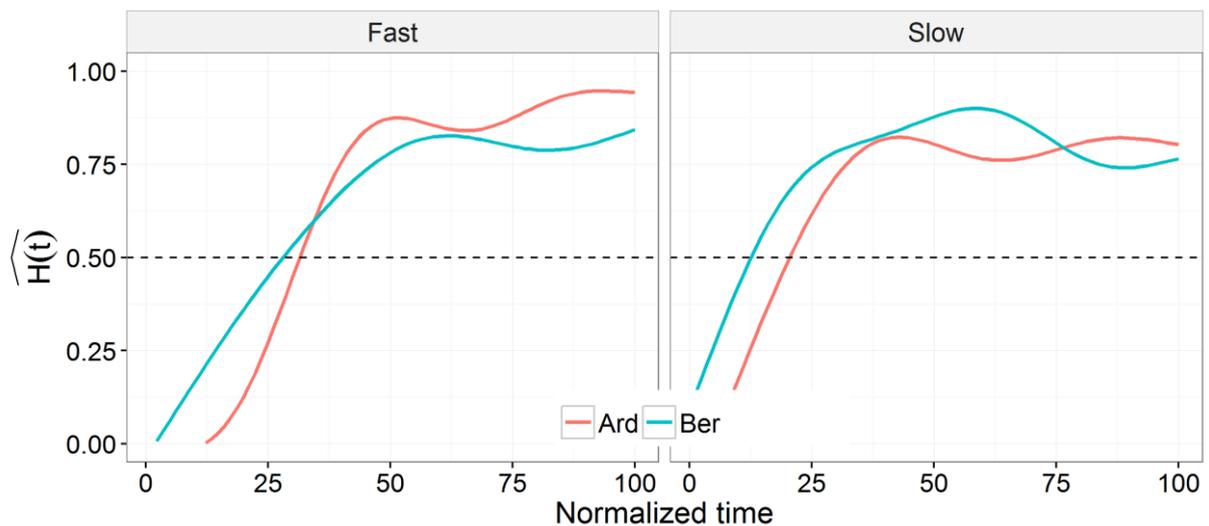

**Figure 4:** Temporal entropy curves for (Ard = Ardenner = hard and (Ber = Berliner = soft) sausages. Time was normalized as a % of total mastication time per measurement (n = 12 for Short (short mastication time), n = 11 for Long (long mastication time), both quadruplicate) TDS data. TDS data from Devezeaux de Lavergne et al. (2015a).

From the results discussed until now we conclude that the temporal entropy curves for different sample categories consumed by different panels follows the same pattern. The uncertainty starts low at the beginning of mastication and rises in the middle of mastication. Towards the point of swallowing the information about the selection of the dominant attribute tends to decrease again except for the case of the two gels $H\sigma L\epsilon T$ and $L\sigma L\epsilon T$ in the first data set, for the two gels HG+HA and LG+HA in the second data set, and for the sausages eaten by the fast eaters in the third data set. We note that the entropy upon swallowing is still higher than 0.5, in line with the fact that there is not one dominant attribute that triggers swallowing. It is likely that a panelist looks for a pattern that being related to a bolus that is safe to swallow and that will not exhibit any unexpected attribute anymore.

We like to discuss the choice of the attributes during the TDS experiment. The texture attributes that are selected are generally based on the results from either a Quantitative Descriptive Analysis (QDA) experiment (Devezeaux de Lavergne et al., 2015b), from previous studies (Devezeaux de Lavergne et al., 2016), or from panel discussions on samples similar to the ones used in the TDS experiment (Devezeaux de Lavergne et al., 2015a). The attributes that are selected for the experiments discussed in this article are chew-down attributes that are perceived during mastication. The attributes to be scored in the TDS experiment are carefully selected to be applicable for the entire set of samples, and the panelists get acquainted with the attributes before the TDS experiment by means of discussing the attributes. We have to realize that a panelist can only actively choose from a limited number of attributes, which, according to Pineau et al. (Pineau et al., 2012), amounts to about 10 attributes. If there are more attributes in the list, not all attributes are used by the panel. The number of attributes in the experiments discussed above was 8 (samples in Figure 2 and 4), or 9 (Figure 3).

Apart from the limitation in number of attributes the time duration for dominance of one attribute should not last too long. This is because if one attribute would be constantly overpowering the others, the uncertainty will stay around 0.5 and the uncertainty about the other attributes, although perceived, will not be accessible (Devezeaux de Lavergne, 2015). This is due to the set-up of TDS where the panelists keep the set of attributes in their mind and need to select the most dominant attribute while consuming and masticating the sample. Such a long-term dominance may be the case in sample HA + HA (cf. Figure 3).

In addition to the set of attributes and the required dynamics in dominance also the amount of training of the panel plays a role in the form of the curve; their experience with sensory experiments and the TDS experimental set-up are important in particular (Meyners, 2011). The panels with the different eating behaviors (cf. Figure 4) where less experienced. Indeed, the slow eaters show less distinct patterns compared to those for the panels scoring the gels (cf. Figure 2 and 3). Interestingly, the panel that performed the experiment with the gels with contrasting layers was more experienced and selected their first dominant attributes earlier than the panelists scoring the gels in Figure 2.

We conclude that the temporal entropy follows a master curve for all samples (gels, gels with mechanical contrast, and sausages) and for two types of eating behavior (i.e. fast or slow eaters). For four out of the twenty samples the entropy does not show a clear decrease upon swallowing, and for one of these the information remains relatively low during the entire mastication process. Nonetheless, the overall shape of a steep rise, a maximum, and a small decline before swallowing seems universal for the types of samples we have investigated. The fact that a master curve is found for many products suggests a possible underlying principle for TDS during mastication for these products. We note that this is not obvious from the TDS data themselves. The possible principle could for instance be the result of underlying fysiological and/or pschychological mechanisms occurring during eating.

In order to further investigate a possible underlying principle we calculate a particular measure for complexity. This complexity measure, $C$ was first introduced by Gershenson (Gershenson, 2007). See also e.g. (Gershenson and Fernández, 2012). Gershenson has given an argument why the complexity can be defined as $C = H(1 - H)$, where $H$ is the entropy, normalized to 1, which is equal to the normalization we have applied in eq.(1). The measure $C$ reflects a combination of randomness and structure, according to Gershenson. Inspired by this work of Gershenson, we plotted in Figure 5 the measure $C$(t) against normalised time for the emulsion filled gels (same samples as in Figure 2).

## 3.2 Complexity and TDS data

We have set out to explore a complexity measure, $C$, which was first introduced by Gershenson (Gershenson, 2007). See also e.g. (Gershenson and Fernández, 2012). Gershenson has given an argument why complexity can be defined as $C = H(1 - H)$, where $H$ is the entropy, normalized to 1, which is equal to the normalization we have applied in eq.(1). The measure $C$ reflects a combination of randomness and structure, according to Gershenson. Inspired by this work of Gershenson, we plotted in Figure 5 the measure $C(t)$ against normalized time for the emulsion filled gels (same samples as in Figure 2). Analogously to the results on Boolean networks, we can note that the rise of $C$ at the beginning and end of mastication may bare an analogy with the rise of $C$ in the critical regime (phase transition) of random Boolean networks, but that this analogy should be investigated to a far greater extent. There may be a connection to phase transitions as reported in other types of systems, such as socio-economic (the stock market), or cognitive systems that gain expertise (Bossomaier et al., 2013).

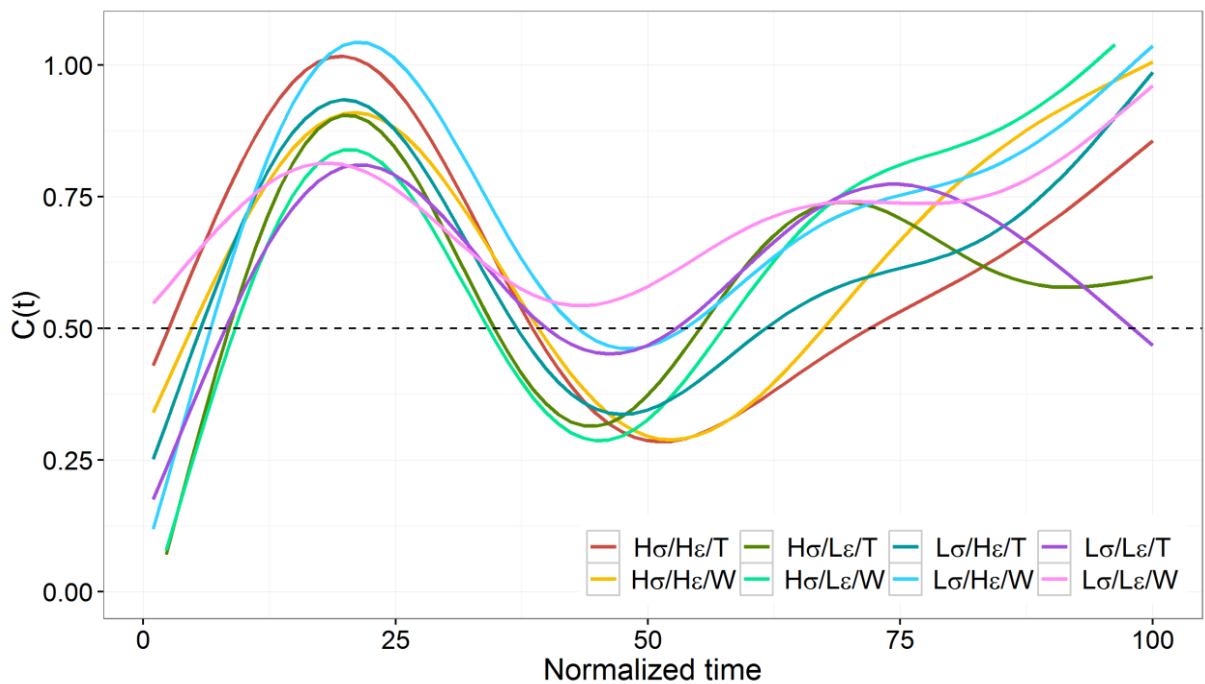

Figure 5: Temporal complexity curves of emulsion-filled gels (with low (L) or high (H) engineered fracture stress ($\sigma$) and strain ($\epsilon$) and WPI (W) or Tween 20 (T) as an emulsifier). Time was normalized as a % of total mastication time per measurement (n = 10, triplicate). TDS data from Devezeaux de Lavergne et al. (2015b).

## 3.3 General Discussion

The above used a quantification of information, or its negative counterpart, entropy, and of the term complexity. The fact that a master curve for entropy arises for many samples is encouraging from a point of view that perhaps there exists a general principle underlying mastication and swallowing of specific types of food products.

We like to mention several aspects on information theory in relation to perception that have not been sorted out in the literature and that are relevant to the above. First of all, during perception, there is a physical interaction between the stimulus and the perceiver, and they can be mutually influenced (Nizami, 2011). In contrast, in the communication system that Shannon describes, the system is in no way altered by the stimulus. Luce (2003) indeed indicates that Shannon himself was skeptical about the use of information theory outside communication engineering. Another point to keep in mind is that people do not behave like static electronic devices, they do not have a "fixed immutable information processing capacity" (Baddeley, 1994). Indeed, with practice, the reaction time to a stimulus levels off. Yet another point is that messages in Shannon's theory of communication often contain redundancy. This helps with encoding and recovering the send message, even in the presence of noise (Shannon, 1948). But, during perception redundancy is not always a fact (Luce, 2003). Another important point of possible criticism would be that, in contrast to information theory, stimuli in psychological experiments have different levels of structural organization and therefore should not be treated as mutually independent statistical events (Aksentijevic and Gibson, 2012). However, there are also proponents of using information theory to sensory science. For example, according to Norwich (2003), perception involves the selection of several choices, therefore making information theory perfectly suitable to apply to problems of perception. And, as another example, Laming (2001) puts forward that information theory can still "provide a 'non-parametric' technique for the investigation of all kinds of systems without the need to understand the machinery, to model the brain without modelling the neural responses."

We note that the measures on entropy and complexity are calculated on the basis of averaging over the responses of all panel members. We have taken this approach for the following reason. If one were to establish these measures per panel member, and obtain accurate data, one would have to perform the experiments many times on the same person. In the current experimental set-up, the same experiment on each panel member has been performed only three or four times to check for intra person variability. Many more repetitions could increase the accuracy of the data per panel member, but could at the same time introduce learning/anticipation into the experiment, which should be avoided as it is another scope of research. Although we calculate the entropy and information for attribute selection for a group of persons, we do believe this still has significance, similar to, for example, the significance of the (average) value of IQ of a group of persons.

Despite the fact that application of information theory to sensory science has not reached overall consensus, we think that our finding of a master curve for TDS data for a variety of food systems provides useful information for the sensory science field, for example in terms of hinting at an underlying physiological and/or psychological principle during mastication and swallowing. This may be as simple as the tendency to only swallow the bolus at the moment it is safe to swallow (as proposed by Hutchinson and Lillford (1988)), and where the perception of the other attributes during chewing would be a side effect. However, the perception of the other attributes during chewing may still be important. In this respect the work by Guevara Erra et al. (2016) on brain function is interesting; they observed a maximum in entropy during normal wakeful states. These authors expressed the hope that their findings could represent a "preliminary attempt at finding organising principles of brain function". In the same way, our results may be a starting point of revealing organising principles during sensory perception.

We also like to mention a couple of aspects of complexity in relation to sensory science that have not been fully sorted out in the literature. First of all, complexity and perceived complexity are used interchangeably while they are not identical; the former is sample dependent, while the latter is both sample and perceiver dependent. Because complexity and perceived complexity are poorly defined (Marcano et al., 2015) it is possible that individuals will allocate different meanings to the term complexity, not only based on their prior experiences, but also in relation to changing contexts (Parr et al., 2011). The issue is further complicated because a stimulus may cause different attributes to emerge, each of them with their own measure for complexity. Thus, often different aspects are studied, which makes an overall comparison impossible. Despite these criticisms, complexity has been linked to hedonic values using the theory of Berlyne (Berlyne, 1970, 1963, 1960). Furthermore, it has been used to underpin many different sensory aspects such as: sensory specific satiety (Weijzen et al., 2008), expected satiating capacity of food products (Marcano et al., 2015), arousal potential and liking (Frøst and Mielby, 2010; Giacalone et al., 2014; Lévy et al., 2006; Mielby et al., 2014, 2013, 2012; Porcherot and Issanchou, 1998; Reverdy et al., 2010; Sulmont-Rossé et al., 2008).

Apart from the critical notes above on using complexity in relation to sensory science we believe that the use of a quantitative measure for complexity as proposed by Gershenson and the observed rise at the beginning and at the end of mastication of this complexity, together with the pattern in complexity, may be observations that could stimulate research on possible underlying mechanisms for mastication and swallowing.

## 4. Conclusions

We find that the entropy, as calculated from the dominant attribute selection by a panel as a function of normalized time (i.e. TDS) before swallowing, follows a master curve for different types of samples, with a maximum in entropy, after which swallowing takes place. The data are in line with the fact that there is not one specific attribute responsible for triggering the swallowing action.

A measure of complexity applied to the TDS data shows also a master curve.

## Acknowledgements

We gratefully acknowledge financial support by TI Food and Nutrition to M. D. and M.S.

# Appendix: Uncertainty, entropy and information

In information theory, or communication theory, like Shannon (Shannon, 1948) called it, a message is transmitted from an information source to a destination with a certain probability. This is schematically depicted in Figure 6. The information source produces a message or sequence of messages as a function of time. The transmitter converts this message to a signal that can be transmitted over the communication channel. The receiver reconstructs the signal. During transmission, noise can be introduced which can cause errors in the signal (Shannon, 1948). An example of this type of communication system can be found in telephony. The information source is the person on one side of the line saying something. The message, the sound waves are converted to an electrical signal that is transmitted over the channel, in this case the telephone wire. At the other end of the line the message is converted back to sound and the other person can hear the message. Usually there is not a one to one correspondence between transmitted versus received signal. The ability to transmit information correctly from the source to the destination is described by the so called capacity of the channel.

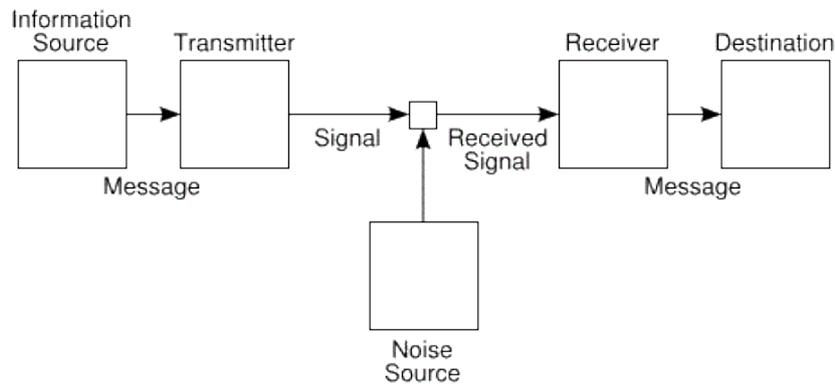

Figure 6: Schematic diagram of a general communication system (adapted from (Shannon, 1948)).

When there is a set of possible events or messages with known probabilities $p_1, p_2, \ldots, p_n$, a measure for the uncertainty regarding which message will be chosen, $H$, is given as:

$$H = -K \sum_{i=1}^{n} p_i \log p_i, \qquad Eq. A1$$

in which $K$ is a positive constant depending on the unit of measure.

An increase in the freedom of choice coincides with an increase in uncertainty that the message that gets selected is actually the correct one (Weaver, 1949). When the probability for one event equals 1, the uncertainty about the event becomes 0, i.e. we are certain about the outcome. When all the probabilities $p_i$ are equal, $p_i = \frac{1}{n}$, the amount of uncertainty increases monotonically with the amount of options $n$ (Shannon, 1948).

The uncertainty regarding which message is chosen, as given by Eq. A 1, is also referred to as entropy. The terms uncertainty, entropy, and information, can be related to one another as follows. According to Rothstein (Rothstein, 1951), in physics one is interested in information obtained from a system by a measurement on the system. The amount of information obtained from a measurement equals the difference between the final information, $I_f$, and initial information, $I_i$, on the system. If the information (about the system) increases, the uncertainty (on the system) decreases. The uncertainty is also referred to as entropy. So, information on a system equals minus the entropy of that system. Or, as Rothstein put it:

"... information obtained from a measurement equals the difference between initial and final entropies of that system". In short: $I_f - I_i = -(S_f - S_i)$ with $S_f$ and $S_i$ the final and initial entropy. As for having another perspective on entropy we can refer to (Brillouin, 1956) " ... The entropy is usually described as measuring the amount of disorder in a physical system. A more precise statement is that entropy measures the lack of organisation about the actual structure of the system".